\documentclass[prd,longbibliography,10pt,aps,twocolumn,amsmath,amssymb,superscriptaddress,nofootinbib,floatfix]{revtex4-1}

\usepackage[english]{babel}

\usepackage[colorlinks=true, linkcolor=violet, citecolor=teal, urlcolor=blue]{hyperref}

\usepackage{cancel}
\usepackage{nicefrac}
\usepackage{graphicx}
\usepackage{float}
\usepackage{slashed}
\usepackage{combelow}
\usepackage{dcolumn}		%decimal point alignment in Tables, use 'd'
\usepackage[dvipsnames]{xcolor}
\usepackage[normalem]{ulem}
\newcommand{\changeq}[3]{{\color{red} \ifmmode\text{\sout{\ensuremath{#1}}}\else\sout{#1}\fi}{\color[rgb]{0.56,0.0,1.0} #2}{\color{blue}[#3]}}
\newcommand{\replaceR}[2]{{{\color{BrickRed}{#1}}{\color{NavyBlue}{\ifmmode\text{\sout{\ensuremath{#2}}}\else\sout{#2}\fi}}}}

\newcommand{\replaceB}[2]{{{\color{RedViolet}{#1}}{\color{BlueViolet}{\ifmmode\text{\sout{\ensuremath{#2}}}\else\sout{#2}\fi}}}}
\newcommand{\vb}[2]{{{\color{RedViolet}{#1}}{\color{BlueViolet}{\ifmmode\text{\sout{\ensuremath{#2}}}\else\sout{#2}\fi}}}}

\newcommand{\replaceC}[2]{{{\color{RedOrange}{#1}}{\color{OliveGreen}{\ifmmode\text{\sout{\ensuremath{#2}}}\else\sout{#2}\fi}}}}

\definecolor{purple}{rgb}{0.8,0,0.6}

\DeclareMathOperator{\Tr}{Tr}

\allowdisplaybreaks
\begin{document}

\title{Chiral and deconfinement thermal transitions at finite quark spin polarization  in lattice QCD simulations}

\author{V.~V.~Braguta} 
\email{vvbraguta@theor.jinr.ru}
\affiliation{Bogoliubov Laboratory of Theoretical Physics, Joint Institute for Nuclear Research, Dubna, 141980 Russia}
\author{M.~N.~Chernodub}
\email{maxim.chernodub@univ-tours.fr}
\affiliation{Institut Denis Poisson, CNRS-UMR 7013, Universit\'e de Tours, 37200 France}
\affiliation{Department of Physics, West University of Timi\cb{s}oara,  Bd.~Vasile P\^arvan 4, Timi\cb{s}oara 300223, Romania}
\author{A.~A.~Roenko}
\email{roenko@theor.jinr.ru}
\affiliation{Bogoliubov Laboratory of Theoretical Physics, Joint Institute for Nuclear Research, Dubna, 141980 Russia}

\date{\today}

%%%%%%%%%%%%%%%%%%%%%%%%%%%%%%%%%%%%%%%%%%%%%%%%%%%%%%%%%%%%%%
\begin{abstract}
We study the effect of finite spin quark density on the chiral and deconfinement thermal crossovers using numerical simulations of lattice QCD with two dynamical light quarks. The finite spin density is introduced by the quark spin potential in the canonical formulation of the spin operator. We show that both chiral and deconfinement temperatures are decreasing functions of the spin potential. We determine the parabolic curvatures of crossover temperatures in a limit of physical quark masses.
\end{abstract}

%%%%%%%%%%%%%%%%%%%%%%%%%%%%%%%%%%%%%%%%%%%%%%%%%%%%%%%%%%%%%%
\maketitle

\section{Introduction}

Properties of spin degrees of freedom in quark-gluon plasma have attracted significant attention~\cite{Becattini:2022zvf}. The interplay of rotation, vorticity, and spins of quarks leads to a series of experimentally observable effects in this strongly interacting medium that emerges in relativistic heavy-ion collisions~\cite{Gao:2020vbh}. On the theoretical side, the investigation of the spin dynamics establishes new links between quantum field theory, thermodynamics, hydrodynamics, and relativistic many-body systems~\cite{Montenegro:2017rbu, Florkowski:2017ruc, Huang:2024ffg}. 

Importantly, the spin effects in quark-gluon plasma can also be accessed experimentally, for example, via measuring the polarization of $\Lambda$ and $\bar{\Lambda}$ hyperons that are produced in non-central heavy-ion collisions~\cite{STAR:2017ckg}. The alignment of spins of these particles with the direction of the angular momentum of the quark-gluon plasma provides experimental evidence for the coupling between macroscopic rotation and microscopic spin degrees of freedom of quarks~\cite{Becattini:2022zvf}. This high-energy phenomenon has deep roots in the celebrated Barnett effect, which was observed more than a century ago: a rotating ferromagnet gets magnetized by rotation via an alignment of intrinsic spins parallel to the rotation axis~\cite{Barnett_1915}.

Thermodynamic properties of vortical quark-gluon plasmas have been subjected to intensive theoretical investigation in various effective analytical models~\cite{Chen:2015hfc, Jiang:2016wvv, Chernodub:2016kxh, Chernodub:2017ref, Wang:2018sur, Chen:2020ath,  Golubtsova:2022ldm, Chen:2022smf, Singha:2024tpo, Jiang:2023zzu, Sun:2023kuu, Chen:2023cjt, Zhao:2022uxc, Yadav:2022qcl, Braga:2022yfe, Mehr:2022tfq, Sadooghi:2021upd, Fujimoto:2021xix, Zhang:2020hha, Chen:2022mhf}. The effect of uniform rotation on the thermal phase transition has also been investigated numerically, both in purely gluonic lattice Yang-Mills theory~\cite{Braguta:2020biu, Braguta:2021jgn, Chernodub:2022veq, Braguta:2023yjn, Braguta:2023kwl, Braguta:2023tqz, Braguta:2023iyx, Braguta:2024zpi} and in lattice QCD with dynamical quarks~\cite{Braguta:2022str, Yang:2023vsw}. A comparison between theoretical and numerical approaches revealed a profound disagreement between predictions of the most theoretical models and the numerical results of first-principle simulations: while the effective infrared models with original, rotation-independent parameters indicate that the pseudocritical crossover temperatures should drop in the vortical plasma, the numerical simulations demonstrate that both deconfinement and chiral crossover temperatures, on the contrary, rise with increased rotation (see discussions in Refs.~\cite{Singha:2024tpo, Morales-Tejera:2025qvh}). A thorough numerical study of Refs.~\cite{Braguta:2022str, Braguta:2024zpi} revealed that this contradiction is a result of the non-perturbative dynamics of gluons, which is difficult to take into account in analytical models.

A window on the properties of the vortical quark-gluon plasma can be opened by investigation of the spin polarization of quarks, which can serve as a spin-sensitive probe of the non-perturbative dynamics of the gluonic component. A finite spin polarization corresponds to the presence of a finite density of spins of quarks in quark-gluon plasma. In a statistical ensemble, a quark spin density can be introduced via a spin potential in exactly the same spirit as a finite baryon density is described via a finite baryon chemical potential. The spin potential shares a distant similarity with the axial and helical chemical potentials~\cite{Ambrus:2019khr, Chernodub:2020yaf}. In our article, we study the effect of the finite quark spin density on the phase diagram of QCD using first-principle Monte Carlo simulations. We concentrate our attention on the chiral and deconfinement thermal crossovers.

It is important to stress the existence of an ambiguity in the very definition of a spin tensor that determines the local spin degrees of freedom. This ambiguity appears as a result of the pseudogauge symmetry~\cite{Becattini:2018duy, Speranza:2020ilk, Fukushima:2020ucl}, which could potentially be fixed by observing that a concrete expression for the spin tensor can be linked to the particularities of the spin interactions in the system~\cite{Buzzegoli:2024mra}. Following the latter article, we will use, in the field-theoretical language, the canonical definition of the spin tensor, which can directly be derived via Noether’s theorem applied to Dirac fermions.

The structure of the paper is as follows. In Section~\ref{sec_spin_potential}, we introduce the spin potential in the Dirac Lagrangian and define the notion of the curvature of the crossover transition with respect to the spin potential. Section~\ref{sec_simulations} is devoted to the description of our numerical setup and the presentation of the results of our Monte Carlo simulations. The last section includes a brief discussion and conclusions.

\section{Finite quark spin density}
\label{sec_spin_potential}

\subsection{Introducing quark spin potential}

Using numerical Monte Carlo simulations, we study the phase diagram of QCD at finite spin density of quarks. The concept of the quark spin density is somewhat similar to a finite baryon density associated with the quark degrees of freedom. The latter can be achieved by introducing the quark chemical potential $\mu_q$, which is conjugated to the quark number density. Analogously, the finite spin density of quarks can effectively be described by the spin potential, $\mu_\Sigma$ which is thermodynamically conjugated with the density of quark spins. A thermodynamic ensemble of quarks characterized by a finite spin potential $\mu_\Sigma \neq 0$ possesses nonzero densities of spin-polarized quarks and anti-quarks such that the total baryon charge vanishes while the spin polarization is nonzero. In other words, in such a configuration, the total numbers of quarks and anti-quarks are equal to each other, implying that the positive baryon charge of quarks is canceled by the negative baryon charge of anti-quarks. However, the spins of quarks and anti-quarks add up so that the total spin in this baryon-neutral configuration is nonzero.

The quark spin density can be described by the following term in the Dirac Lagrangian:
\begin{align}
    \delta_{\Sigma}\, {\mathcal L}_q = \mu_{\alpha,\mu\nu} \, \overline{\psi} {\cal S}^{\alpha,\mu\nu} \psi\,,
    \label{eq_L_Spin}
\end{align}
where we introduced the relativistic spin density matrix,
\begin{align}
    {\cal S}^{\alpha,\mu\nu} = \frac{1}{2} \bigl\{\gamma^\alpha, \Sigma^{\mu\nu}\bigr\}\,,
    \qquad
    \Sigma^{\mu\nu} = \frac{i}{4} \bigl[\gamma^\mu, \gamma^\nu\bigr]\,,
    \label{eq_S_tensor}
\end{align}
which plays the role of the canonical relativistic spin current operator for spin-1/2 Dirac fermions. The associated background relativistic spin field $\mu_{\alpha,\mu\nu}$ introduces a finite spin density and a finite spin current in the system.

Without losing generality, we consider the spins of quarks polarized along the $z$ axis in a certain inertial reference frame. We choose the background relativistic spin field as follows: 
\begin{align}
	\mu_{\alpha,\mu\nu} = \frac{\mu_\Sigma}{2} \delta_{\alpha 0} \bigl(\delta_{\mu 1} \delta_{\nu 2} - \delta_{\nu 1} \delta_{\mu 2} \bigr)\,,
    \label{eq_mu_S_tensor}
\end{align}
with all other components vanishing. We use the Minkowski metric $g = {\rm diag} (1,-1,-1,-1)$.

The Lagrangian~\eqref{eq_L_Spin} introduces a finite quark-antiquark density with their spins polarized along the $z$ direction, with the global spin polarization controlled by the spin potential $\mu_\Sigma$. With Eqs.~\eqref{eq_S_tensor} and \eqref{eq_mu_S_tensor} in place, the simplified Lagrangian~\eqref{eq_L_Spin} gets the following form:
\begin{align}
    \delta_{\Sigma} {\mathcal L}_q = \mu_\Sigma \overline{\psi} \gamma^0 \Sigma^{12} \psi\,.
    \label{eq_L_psi}
\end{align} 
In the Dirac representation, the matrix in the above equation has a diagonal form: 
\begin{align}
	\gamma^0 \Sigma^{12} & \equiv {\cal S}^{0,12} \equiv -  {\cal S}^{0,21} = \frac{1}{2} \gamma^3 \gamma^5 \nonumber \\ 
    & = {\rm diag} \bigl( +\tfrac{1}{2}, -\tfrac{1}{2}, -\tfrac{1}{2}, +\tfrac{1}{2} \bigr)\,.
    \label{eq_equiv_gamma}
\end{align}
The quantity $\mu_\Sigma$ is called ``spin potential'' as it serves as a potential energy for spin coordinate as and also has a certain similarity with a chemical potential\footnote{Following Ref.~\cite{Becattini:2022zvf}, we call $\mu_\Sigma$ a spin potential and not a spin {\it chemical} potential because the spin is not a conserved quantity.}: one needs to use the $\delta E_{\uparrow} = +\mu_\Sigma/2$ amount of energy to add a spin-$1/2$ state that points up along the $z$ axis and one requires investing $\delta E_{\downarrow} = - \mu_\Sigma/2$ energy to add a spin-$1/2$ state that points down.

Thus, the spin potential $\mu_\Sigma$ of relativistic fermions is a thermodynamic quantity that describes the energy difference between adding a spin-$1/2$ particle with one orientation of a spin versus the other to a system of relativistic fermions, $\mu_\Sigma = \delta E_{\uparrow} - \delta E_{\downarrow}$. In other words, it expresses a tendency to favor one spin state over the other, and it is a way to account for the spin degree of freedom in the system. 

By construction, the notion of the spin potential $\mu_\Sigma$ is somewhat similar to the notion of the angular velocity $\Omega$ that characterizes a uniformly rotating system. Indeed, in a reference frame that co-rotates together with the system with angular velocity $\boldsymbol{\Omega} = \Omega\, {\bf e}_z$ about the $z$ axis, the rotation affects the quark Lagrangian by shifting it with the following term (see, for example, a derivation in Ref.~\cite{Chen:2015hfc} or Ref.~\cite{Chernodub:2016kxh}):
\begin{align}
	\delta_{\Omega} {\mathcal L}_q = 
    \Omega \overline{\psi} \bigl[i (-x \partial_y + y \partial_x) + \gamma^0 \Sigma^{12} \bigr]\psi\,,
    \label{eq_L_psi_Omega}
\end{align}
where the first term represents an orbital momentum of the quarks while the second term corresponds to an effect of global uniform rotation on the quark spin. Identifying for a brief moment $\Omega = \mu_\Sigma$, we find that the spin-polarization term for the orbital motion in Eq.~\eqref{eq_L_psi_Omega} coincides precisely with the Lagrangian term that describes the spin polarization in the non-rotating medium~\eqref{eq_L_psi}. 

The orbital motion of the quarks is, evidently, absent in our case because we aim to study a spin-polarized but non-rotating medium~\eqref{eq_L_psi}. Our approach has, thus, an advantage in that it can be studied in the thermodynamic limit since the spin polarization itself does not cause the causality problem that requires restricting the rotating system to be located entirely within a light cylinder~\cite{Davies:1996ks}.

Before continuing further, it is important to stress that we consider the spin polarization imposed only on quark fields. Spins of gluons are not directly affected by our background. However, it is not excluded (and, moreover, it is quite plausible) that the polarized quark medium does affect the gluon spin polarization indirectly, via quark loops, as the spins of quarks transfer to the spins of gluons and vice versa.

\subsection{Curvature of the crossover transition}

In the imaginary time formalism, the quark spin potential becomes an imaginary quantity after the Wick rotation, similarly to the baryon chemical potential. Due to the notorious sign problem, Monte Carlo simulations with a real spin potential in the Wick-rotated Euclidean field theory are impossible. Therefore, following our experience with the quark chemical potential, we proceed by introducing the imaginary spin potential, 
\begin{align}
	\mu_\Sigma = i \mu_\Sigma^{\rm I}\,,
\end{align}
and then use an analytical continuation from imaginary to real values of the spin potential for our results obtained in Monte Carlo simulations. 

The procedure of the analytical continuation is justified at small values of the spin potential $\mu_\Sigma^{\rm I}$. In our paper, we aim to find the curvature $\kappa_\Sigma$ of the deconfinement ($\ell = L$) and chiral ($\ell = \psi$) crossover transitions that determine the behavior of the corresponding 
pseudocritical temperatures,
\begin{align}
	\frac{T^\ell_c(\mu_\Sigma^{\rm I})}{T^\ell_{c}(0)} =  1 + \kappa^\ell_\Sigma \Bigl(\frac{\mu^{\rm I}_\Sigma}{T_c(0)}\Bigr)^2 + \dots\,,
    \qquad {[\ell = L, \psi]}\,,
    \label{eq_T_muI_S}
\end{align}
where higher-order terms are shown by the ellipsis. 

Analytically continuing the parabolic function~\eqref{eq_T_muI_S} back to the real-valued quark spin potential, $\mu_\Sigma^{\rm I} \to i \mu_\Sigma$, we get the behavior of the pseudocritical temperatures as a function of $\mu_\Sigma$:
\begin{align}
	\frac{T_c^{\ell}(\mu_\Sigma)}{T_{c}(0)} =  1 - \kappa_\Sigma^{\ell} \Bigl(\frac{\mu_\Sigma}{T_c(0)}\Bigr)^2 + \dots\,, 
    \qquad {[\ell = L, \psi]}\,.
\label{eq_Tc_mu_S}
\end{align}
The spin curvature $\kappa_\Sigma$ describes how the presence of small quark density affects the deconfining crossover: a positive (negative) $\kappa_\Sigma$ would imply that the pseudocritical temperature diminishes (grows) as the spin density increases. In this article, we compute numerically the thermal ``spin'' curvature $\kappa_\Sigma$ of the deconfining crossover transition~\eqref{eq_Tc_mu_S}. Surprisingly, our simulations demonstrate that the parabolic regime of Eqs.~\eqref{eq_T_muI_S} and \eqref{eq_Tc_mu_S} extends well beyond the region of small spin potentials.

At the end of this section, we notice a curious analogy that emerges, at zero temperature, between the systems described, on one side, by the imaginary quark spin potential $\mu^{\rm I}_\Sigma$ in the Euclidean spacetime  and, on the other side, the axial (chiral) chemical potential, $\mu_A$ both in Euclidean and Minkowski spacetimes (we remind the reader that a finite axial density does not lead to the sign problem). According to Eq.~\eqref{eq_equiv_gamma}, a finite spin polarization along a spatial direction is proportional to the axial current density along the same direction, $j_A^{3} =  \overline{\psi} \gamma^{3} \gamma^5 \psi \equiv 2 \overline{\psi} \gamma^0 \Sigma^{12} \psi$, where we took the spatial direction $\mu = 3$ for definiteness. Therefore, a spin potential generates an axial current along the corresponding spatial direction. At zero temperature Euclidean field theory, the spatial directions and the imaginary time direction are equivalent. Therefore, the spin polarization generated by the spin potential $\mu^{\rm I}_\Sigma$ is equivalent to the axial density produced by the axial chemical potential $\mu_A = \mu^{\rm I}_\Sigma/2$.

QCD at finite axial chemical potential $\mu_A$ was studied intensively both theoretically (see, for instance, papers~\cite{Chernodub:2011fr, Braguta:2016aov, Ruggieri:2016ejz, Khunjua:2018jmn, Ruggieri:2020qtq, Yang:2020ykt}) and within lattice simulations~\cite{Braguta:2015zta, Braguta:2015owi, Astrakhantsev:2019wnp}. Among the results related to this paper, one could mention the chiral catalysis phenomenon~\cite{Braguta:2016aov}, which enhances the chiral condensate in the presence of a non-zero axial (chiral) density generated by a finite axial chemical potential $\mu_A$. In addition, the QCD string tension is enhanced as the chiral density gets larger~\cite{Astrakhantsev:2019wnp}. As a result, the pseudocritical temperatures of chiral and deconfinement crossovers are increased as the chiral chemical potential gets larger~\cite{Braguta:2015zta, Braguta:2015owi}. Because of the mentioned isotropy of the Euclidean spacetime at zero temperature, the same effect also holds a finite spin potential $\mu_\Sigma^{\rm I}$. Since the increasing spin potential enhances the chiral condensate and the confinement property, it works oppositely to the effect of thermal fluctuations that tend to diminish both the chiral condensate and decrease the tension of the confining string. This observation suggests that the transition temperature rises with the increase of $\mu_\Sigma^{\rm I}$, thus implying $\kappa_\Sigma > 0$ in Eq.~\eqref{eq_T_muI_S}. Below, we calculate $\kappa_\Sigma$ numerically using first-principle lattice simulations.

\section{Simulations at finite spin density}
\label{sec_simulations}

\subsection{Lattice setup}

The incorporation of a quark spin potential into the lattice QCD action shares some similarity with imposing a uniform rotational background, albeit with two significant simplifications.

First of all, we stress that the quark spin potential, as it follows from the name of this quantity, is applied explicitly only to the quark degrees of freedom. Therefore, the gluons are not subjected {\it directly} to rotation. In notations of Ref.~\cite{Braguta:2022str}, the angular velocity of gluon fields is zero, $\Omega_G = 0$. 

Moreover, in order to impose the quark spin potential, we should not rotate quarks in the orbital sense. In continuum notations, the Euclidean fermionic action at finite (imaginary) spin potential $\mu_\Sigma^{\rm I}$ has the following form:
\begin{align}
S_F = & \int d^4x \ \bar{\psi} \Bigl[ \gamma^x D_x + \gamma^y D_y + \gamma^z D_z \nonumber \\
& + \gamma^\tau \bigl( D_\tau + i \mu_\Sigma^{\rm I} \Sigma^{12} \bigr) + m \Bigr] \psi\,,
\label{eq_SF2}
\end{align}
where $D_\mu$ is the covariant derivative and $\Sigma^{\mu\nu}$ is the spin matrix defined in Eq.~\eqref{eq_S_tensor}.

In action~\eqref{eq_SF2}, the spin polarization is taken into account by the $\Sigma^{12}$ term, while the orbital rotation may be encoded in the change in the Dirac gamma matrices~\cite{Yamamoto:2013zwa}. Since the orbital term is not needed for our aims, the gamma matrices remain unmodified: $\gamma^x = \gamma^1$, $\gamma^y = \gamma^2$, $\gamma^z = \gamma^3$, and $\gamma^\tau = \gamma^4$, so that the quarks do not perform the orbital motion. Thus, in order to incorporate the spin polarization, we need only the spin-rotation coupling term $i \gamma^\tau \Omega \Sigma^{12}/2$, where we take $\Omega \to \mu_\Sigma$. Notice that we omitted above an index of the chemical potential, $\mu_{s} \equiv \mu_{s;z}$ since the spin is assumed to be induced in the $z$ direction.

Technically, the numerical simulations of lattice QCD at finite quark spin density have noticeable simplifications compared to the simulations of quark-gluon matter in the rotating frame: the system is homogeneous, and there is no need to impose non-trivial boundary conditions in $x,y$-direction
in order to enforce the causality property. Thus, the simulations at finite quark spin density are rather similar to QCD at finite baryon density or chiral chemical potential, and, at the same time, they are similar to the simulations of the rotating QCD in which the rotating gluonic and rotating orbital quark contributions are omitted. In order to simplify the technical implementation of the spin potential and make it possible to compare the results of vortical and quark-spin-polarized plasmas, we used a modification of the same lattice action as in our ongoing study of rotating QCD~\cite{Braguta:2022str}, with vortical gluon and orbital quark parts excluded.

We discretize the gluon part of the action using the renormalization-group improved (Iwasaki) lattice action~\cite{Iwasaki:1985we}, which is unaffected by spin density:
\begin{align}
S_{G} = \beta  \sum_{x} \Big( c_0\sum_{\mu < \nu} W^{1\times1}_{\mu\nu}  + c_1 \sum_{\mu \neq \nu} W^{1\times2}_{\mu\nu} \Big)\,,
\label{eq_action_lat_imp_G}
\end{align}
with the lattice couplings $\beta = 6/g^2$,  $c_0 = 1 - 8 c_1$, and $c_1 = -0.331$. The gauge field enters via
\begin{align}
    W^{1\times1}_{\mu\nu}(x) & = 1 - \frac{1}{3} {\rm Re} \Tr \, \overline{U}_{\mu\nu}(x)\,, \\
    W^{1\times2}_{\mu\nu}(x) & = 1 - \frac{1}{3} {\rm Re} \Tr \, R_{\mu\nu}(x)\,, 
\end{align}
where ${\overline U}_{\mu\nu}(x)$ denotes the clover-type average of four plaquettes and $R_{\mu\nu}(x)$ represents a rectangular loop~\cite{Braguta:2021jgn}.

The gauge action is supplemented by the $N_f=2$ clover-improved action for Wilson fermions~\cite{Sheikholeslami:1985ij}, which now incorporates the imaginary spin potential $\mu^{\rm I}_\Sigma$:
\begin{align}
    S_F = \sum_{f=u,d}\sum_{x_1, x_2} \bar\psi^{f}(x_1) M_{x_1,x_2} \psi^{f}(x_2)\,,
    \label{eq_action_F}
\end{align}
with the matrix
\begin{align}
	    M_{x_1, x_2} & {} = \delta_{x_1,x_2} - {} \nonumber \\
        &{} - \kappa \bigg[
        \sum_{\mu = x,y,z} \Big( (1-\gamma^\mu) T_{\mu+} + (1+\gamma^\mu) T_{\mu-} \Big) + {} \nonumber \\
        &{} + (1-\gamma^\tau)\, \exp\Big( {i a \mu^{\rm I}_\Sigma \Sigma^{12} } \Big) T_{\tau+} + {} \nonumber \\
        &{} + (1+\gamma^\tau)\, \exp\Big( {-i a \mu^{\rm I}_\Sigma \Sigma^{12} } \Big) T_{\tau-}
        \bigg] - {} \nonumber \\
        &{} - \delta_{x_1,x_2} c_{SW} \kappa \sum_{\mu<\nu} \sigma_{\mu\nu}F_{\mu\nu}\,,
\end{align}
where $\kappa = 1/(8+2am)$, $T_{\mu+} = U_\mu(x_1) \delta_{x_1+\mu, x_2}$, $T_{\mu-} = U_\mu^\dagger(x_1) \delta_{x_1-\mu, x_2}$ and $F_{\mu\nu} = (\bar{U}_{\mu\nu} - \bar{U}^\dagger_{\mu\nu})/8i$. For the clover coefficient, we adopt the mean-field value $c_{SW} =  (1-W^{1\times1})^{-3/4} = (1 - 0.8412/\beta)^{-3/4}$ following Refs.~\cite{Maezawa:2007fc, Ejiri:2009hq} and then substitute a one-loop result for the plaquette~\cite{Iwasaki:1985we}.

For this lattice action, the masses of light mesons were calculated for a wide range of simulation parameters in Refs.~\cite{CP-PACS:2000phc, CP-PACS:2000aio, CP-PACS:2001hxw, CP-PACS:2001vqx}. We reanalyze that data to restore the lines of constant physics, and our interpolation results are consistent with previous studies~\cite{CP-PACS:2001hxw, Maezawa:2007fc} within systematic uncertainties. Simulations are performed on lattices of the size $4\times 16^3$, $5\times 20^3$, $6\times 24^3$ for meson mass ratios $m_{\rm PS}/m_{\rm V} = 0.60, \dots, 0.85$. Note that in rotating QCD, we found that the fermionic and gluonic degrees of freedom have opposite effects on the pseudocritical temperature~\cite{Braguta:2022str}. Therefore, the dependence of the results on the pion mass -- which controls the dynamical properties of fermions -- is of particular interest.

\subsection{Observables}

At zero baryon density, quark-gluon matter can exist in two distinct phases. The low-temperature phase possesses the color confinement property and exhibits the chiral symmetry breaking, implying that the physical degrees of freedom are massive, colorless hadronic states. The high-temperature phase is the quark-gluon plasma, in which the quarks and gluons are deconfined and the chiral symmetry is restored. At real physical quark masses, the transition between these phases is a smooth crossover~\cite{Aoki:2006we}, which possesses no thermodynamic singularity. In our paper, we are interested in both the confining and chiral properties of quark-spin-polarized QCD. 

In the absence of dynamical quarks, QCD is reduced to Yang-Mills theory, where the transition between the confining and deconfining phases is a true thermodynamic phase transition of the first order. The phases are distinguished by the expectation value of the Polyakov loop, which is an order parameter of the deconfinement phase:
\begin{align} 
    L & = \left\langle |L_{\rm bulk}| \right\rangle\,, 
    \qquad 
    L_{\rm bulk} =\frac{1}{V} \int d^3 r \, L({\boldsymbol r})\,,
    \label{eq_L_continuum_bulk}
     \\
    L(\boldsymbol r) & {} = {\mathrm {Tr}}\, {\mathcal P} \exp \biggl(\oint_{0}^{1/T} d \tau A_4( {\tau}, {\boldsymbol r}) \biggr) \,,
    \label{eq_L_continuum}
\end{align}
where ${\mathcal P}$ is a path ordering operator. Notice that both confining and deconfining regimes represent homogeneous phases so that the expectation value of the local Polyakov loop  $\langle L({\boldsymbol{r}})\rangle $ does not depend on the spatial point ${\boldsymbol r}$. The system is expected to maintain the spatial homogeneity also at small values of the spin potential.

In the low-temperature confinement phase, the expectation value of the Polyakov loop~\eqref{eq_L_continuum} vanishes, $L = 0$, indicating that the free energy of a single heavy quark, $F_{Q} = - T \ln |L|$ diverges ($F_{Q} \to \infty$). Therefore, free isolated quarks do not exist in the low-temperature phase. The vanishing expectation value of the Polyakov loop implies that the center ${\mathbb Z}_3$ global symmetry is respected by the ground state $L = 0$, which is invariant under the global transformations $L \to e^{i 2 \pi n/3}  L$, with $n=0,1,2$.

In the high-temperature phase, the Polyakov loop does not vanish, $L \neq 0$, and the center ${\mathbb Z}_3$ symmetry is broken by a nonzero value of the Polyakov loop. Thus, the free energy of a single quark is a finite quantity, $F_{Q} < \infty$, and quarks can propagate without being confined to colorless bound states. 

In the presence of dynamical matter fields, the center ${\mathbb Z}_3$ symmetry is no more respected. Therefore, the Polyakov loop becomes only an approximate order parameter of the deconfining crossover transition of QCD. Despite this apparent inconsistency, we can still use the inflection point of the Polyakov loop expectation value as a function of temperature to determine the crossover position. Alternatively, we can associate this pseudocritical temperature with the position of the peak of the susceptibility of the bulk Polyakov loop~\eqref{eq_L_continuum_bulk}, 
\begin{equation}
    \chi_L = N_s^3\big( \langle |L_{\rm bulk}|^2 \rangle  - \langle |L_{\rm bulk}|\rangle^2 \big)\,,
    \label{eq_Polyakov_Susceptibility}
\end{equation}
where $N_s$ is a lattice size in spatial directions. While both these formal definitions give comparable values, given a substantial width of the crossover in the thermodynamic limit, we will use the maximum of the susceptibility~\eqref{eq_Polyakov_Susceptibility} in our determination of the pseudocritical temperature of the deconfining crossover. 
It is worth to note that as was shown in paper~\cite{Clarke:2020htu} the quark mass dependence of the Polyakov loop is sensitive to the chiral transition. However, this property of the Polyakov loop will not be used in our paper.

On the lattice, the local Polyakov loop~\eqref{eq_L_continuum} has the following definition:
\begin{equation}
    L(\boldsymbol r) = \left\langle \frac{1}{3} \Tr \left[ \prod_{\tau = 0}^{N_t - 1} U_4(\tau, \boldsymbol r) \right] \right\rangle\,,
    \label{eq_polyakov_loop_lattice}
\end{equation}
where $U_4(\tau, \boldsymbol r)$ is the link variable in the temporal direction. The bulk expectation value is given by Eq.~\eqref{eq_L_continuum_bulk}, where the integral is replaced by a sum over the lattice sites.

In QCD with dynamical quarks, the deconfinement crossover is accompanied by the restoration of chiral symmetry. For the chiral crossover, we determine the pseudocritical temperature using the (disconnected) chiral susceptibility:
\begin{equation}
    \chi_{\bar \psi \psi}^{\rm disc} = \frac{N_f T}{V} \left[\left\langle \text{Tr} (M^{-1})^2 \right\rangle - \left\langle \text{Tr} (M^{-1}) \right\rangle^2 \right]\,,
\end{equation}
which has a peak at the pseudocritical temperature.

In our work, we use non-renormalized susceptibilities of the chiral condensate and Polyakov loop. 
This approach might shift the peak positions of the corresponding susceptibilities as compared to the renormalized ones~\cite{Bazavov:2016uvm}. Notice, however, that in our paper we mainly focus on the ratios of the pseudocritical temperatures $T_c(\mu_{\Sigma})/T_c(0)$. We expect that possible shifts of the peak positions are canceled in the ratios. Notice also that in papers \cite{Braguta:2015zta, Braguta:2015owi} the influence of the axial chemical potential on the ultraviolet divergencies in the chiral condensate and Polyakov loop was studied. It was found that axial chemical potential softens divergency in the chiral condensate considerably and does not lead to additional divergencies in the Polyakov loop. Since the axial chemical potential and the spin potential are different components of the same current we believe that the same is true for the spin potential. Thus one can expect that if one uses renormalized chiral condensate and Polyakov loop possible change of our results for curvatures will be small. Finally our results reveal moderate cutoff dependence (see Fig.~\ref{fig_Tc_Nts}) which also supports the possibility to use non-renormalized susceptibilities to study the ratios of the pseudocritical temperatures.

\subsection{Results}

\begin{figure}[t]
\centering
\includegraphics[width=0.98\linewidth]{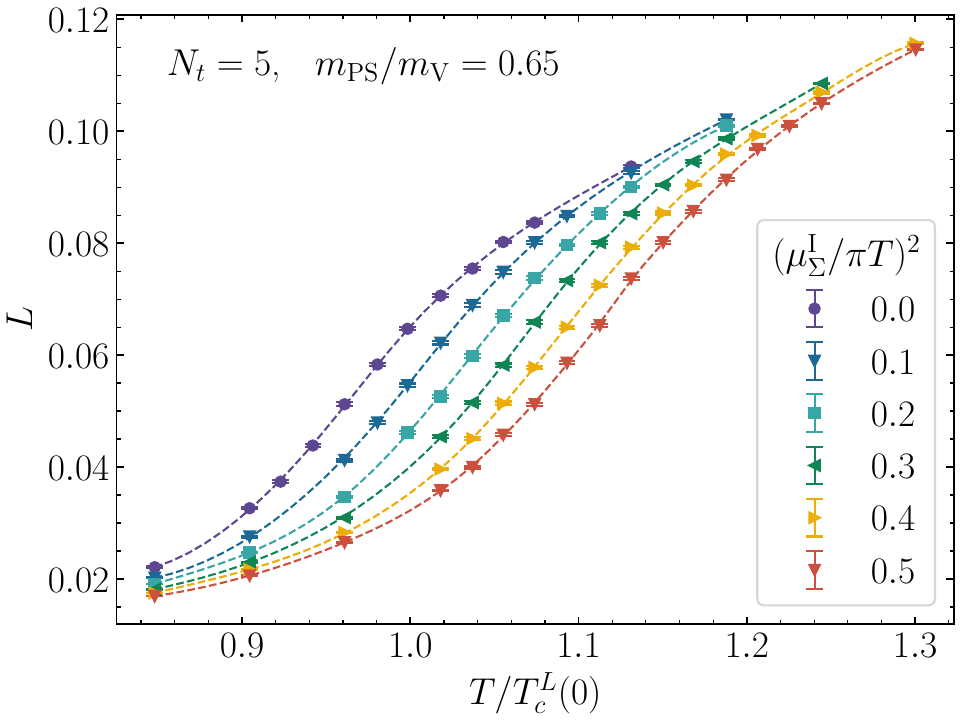} 
\caption{Expectation value of the Polyakov loop as a function of temperature $T$ for various values of the imaginary spin potential $\mu_\Sigma^{\rm I}$. The calculations were performed at the temporal extension of the lattice $N_t = 5$ at the meson ratio $m_{\rm PS}/m_{\rm V} = 0.65$. Temperature and spin potential $\mu_\Sigma^{\rm I}$ are given in units of the deconfinement crossover temperature at vanishing spin density, $T_c^L(0)$. The dotted lines are drawn to guide the eye.
}
\label{fig_Polyakov_Loop}
\end{figure} 

We show the expectation value of the bulk Polyakov loop $L$ as a function of temperature in Fig.~\ref{fig_Polyakov_Loop}. The data are presented for a single value of the pseudoscalar meson mass, encoded in the ratio $m_{\rm PS}/m_{\rm V} = 0.65$, and for various values of the imaginary spin potential $\mu_\Sigma^{\rm I}$ at the temporal extension of the lattice $N_t = 5$. 

An increasing imaginary spin potential leads to a decrease of the Polyakov loop at a fixed temperature while keeping the low-temperature and high-temperature asymptotics largely intact. This behavior implies that at a finite spin density, the crossover shifts to the higher temperatures as compared to the case with vanishing spin density.

\begin{figure}[t]
\centering
\includegraphics[width=0.98\linewidth]{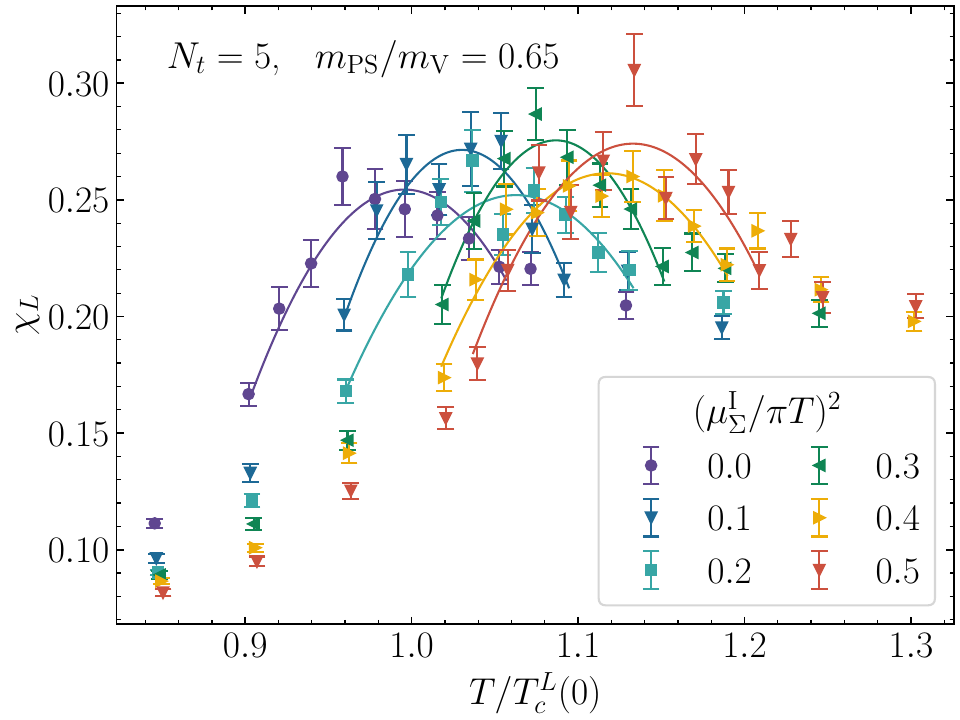} 
\\[1em]
\includegraphics[width=0.98\linewidth]{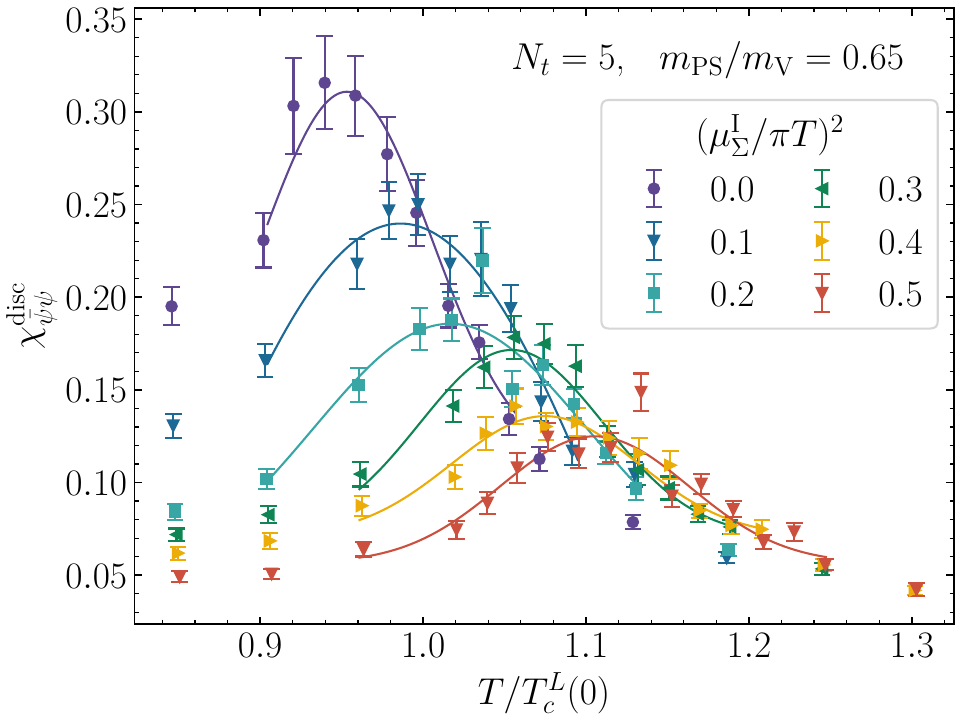} 
\caption{The same as in Fig.~\ref{fig_Polyakov_Loop} but for the susceptibilities of (top) the Polyakov loop and (bottom) the chiral condensate. The solid lines correspond to the best fits of the numerical data by a Gaussian function.
}
\label{fig_susceptibilities}
\end{figure} 

The crossover temperature for the deconfinement can be associated with the maximum of the Polyakov loop susceptibility~\eqref{eq_Polyakov_Susceptibility} as shown in Fig.~\ref{fig_susceptibilities}(top). Similarly, the chiral crossover point can be obtained from the maximum of the (disconnected) susceptibility of the chiral condensate, Fig.~\ref{fig_susceptibilities}(bottom).

\begin{figure}[t]
\centering
\includegraphics[width=0.98\linewidth]{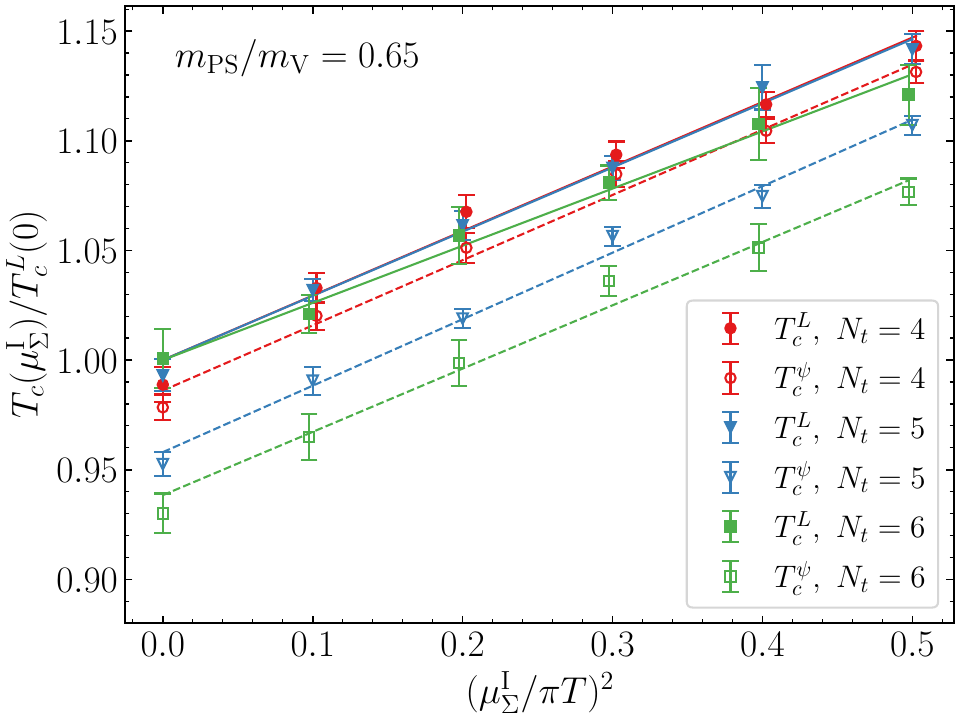} 
\\[1em]
\includegraphics[width=0.98\linewidth]{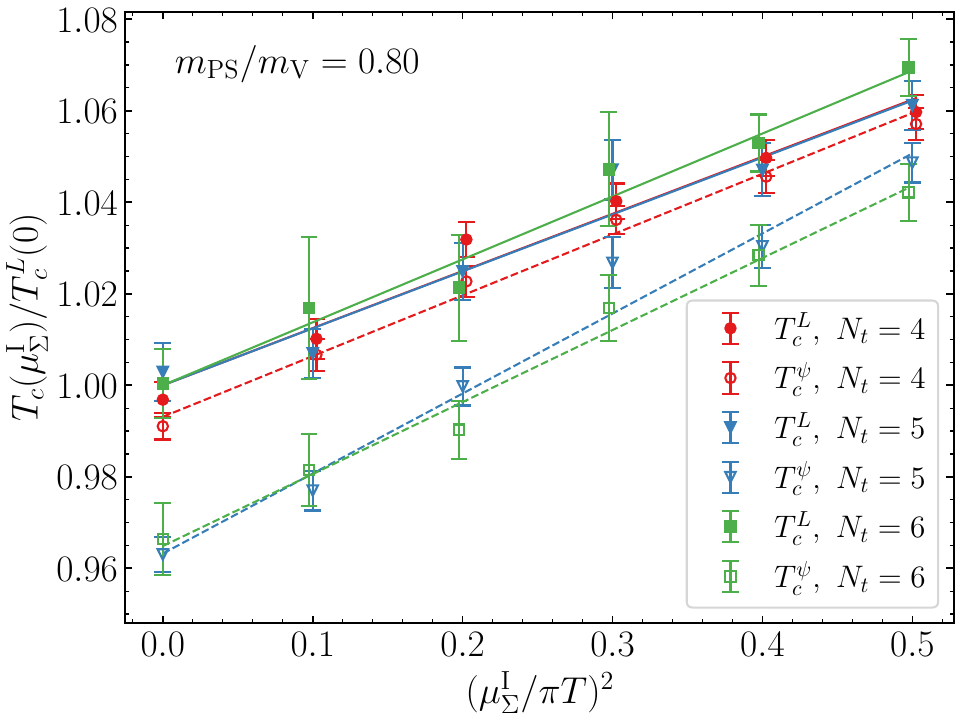} 
\caption{Pseudocritical temperatures $T_c$ of the deconfinement (filled markers) and chiral (empty markers) crossovers at the meson ratios (top) $m_{\rm PS}/m_{\rm V} = 0.65$ and (bottom) $m_{\rm PS}/m_{\rm V} = 0.8$ as a function of the squared imaginary spin potential $\mu^{\rm I}_\Sigma$ (normalized by $\pi T$) for the temporal extensions $N_t = 4,5,6$. The solid lines represent the linear fits given by Eq.~\eqref{eq_T_muI_S_fit}. For convenience, the data are normalized by the fit coefficient~$T_c^L(0)$ (i.e., by the pseudocritical crossover temperature of the deconfinement crossover in the absence of the spin polarization).
}
\label{fig_Tc_Nts}
\end{figure} 

Both pseudocritical temperatures are shown in Fig.~\ref{fig_Tc_Nts} as functions of the imaginary spin potential $\mu^{\rm I}_\Sigma$, for two meson mass ratios $m_{\rm PS}/m_{\rm V}$ and various temporal extensions of the lattice $N_t = 4,5,6$. At a fixed temperature $T = 1/(N_t a)$, as the number of sites $N_t$ in the imaginary temporal direction increases, the lattice spacing decreases, thus allowing us to test the scaling of the data in the ultraviolet limit. The data in Fig.~\ref{fig_Tc_Nts} indicate moderate dependence of the transition temperature on the ultraviolet lattice cutoff played by the lattice spacing $a$. Moreover, it shows a quadratic behavior of the pseudocritical temperature on the imaginary spin potential in consistency with the expected behavior~\eqref{eq_T_muI_S}. 

In order to find the curvatures of the deconfinement and chiral crossover transitions, we fit the data for the pseudocritical temperatures by the following function of the imaginary spin potential $\mu^{\rm I}_\Sigma$:
\begin{align}
	T^\ell_c(\mu_\Sigma^{\rm I}) = T^\ell_{c}(0) \left[ 1 + \kappa^\ell_\Sigma \Bigl(\frac{\mu^{\rm I}_\Sigma}{T} \Bigr)^2 \right] \,,
    \label{eq_T_muI_S_fit}
\end{align}
where the index $\ell$ labels the deconfinement $(\ell = L)$ and chiral $(\ell = \psi)$ pseudocritical temperatures. For each crossover $\ell = L, \psi$, the fitting function~\eqref{eq_T_muI_S_fit} has two fitting parameters that fix the crossover temperature $T^\ell_{c}(0)$ at a vanishing imaginary spin potential $\mu^{\rm I}_\Sigma$, and the curvature $\kappa^\ell_\Sigma$ of the transition at small values of $\mu^{\rm I}_\Sigma$. For convenience, the data for all pseudocritical crossover temperatures in Fig.~\ref{fig_Tc_Nts} are normalized by the temperature $T_c^L(0)$ of the deconfinement crossover at vanishing spin density. Notice that up to inessential quartic terms, Eq.~\eqref{eq_T_muI_S_fit} can also be rewritten in the more familiar form~\eqref{eq_T_muI_S}.
The slope of the fitting functions in Fig.~\ref{fig_Tc_Nts}, parametrized by the curvature $\kappa_\Sigma^\ell$, has a minor dependence on the lattice spacing. Note that this dependence on the ultraviolet cutoff is much less pronounced than for the crossover temperatures itself.

\begin{figure}[t]
\centering
\includegraphics[width=0.98\linewidth]{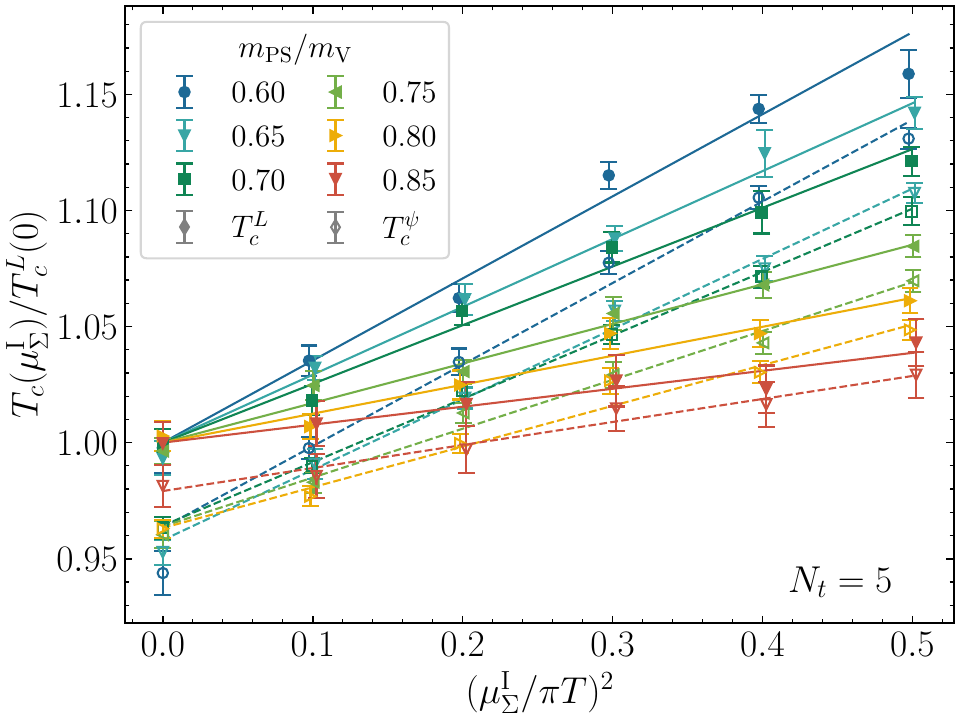} 
\caption{Pseudocritical temperatures $T_c^L$ and $T_c^\psi$ as a function of the (normalized) squared spin potential $\mu_\Sigma^{\rm I}$ for various ratios of the meson masses $m_{\rm PS}/m_{\rm V}$ at the $N_t = 5$ lattice. The solid (dotted) lines represent the linear fits for deconfinement (chiral) pseudocritical temperatures given by Eq.~\eqref{eq_T_muI_S_fit}. The data are normalized by the pseudocritical crossover temperature of the deconfinement crossover $T_c^L(0)$.
}
\label{fig_Tc_muI}
\end{figure}

The dependence of the pseudocritical temperature on the imaginary spin potential for all available ratios of the meson masses $m_{\rm PS}/m_{\rm V}$ is summarized in Fig.~\ref{fig_Tc_muI}. The data point out that at each fixed value of the ratio, the transition temperature is a parabolic function of the imaginary spin potential. 

\begin{figure}[t]
\centering
\includegraphics[width=0.98\linewidth]{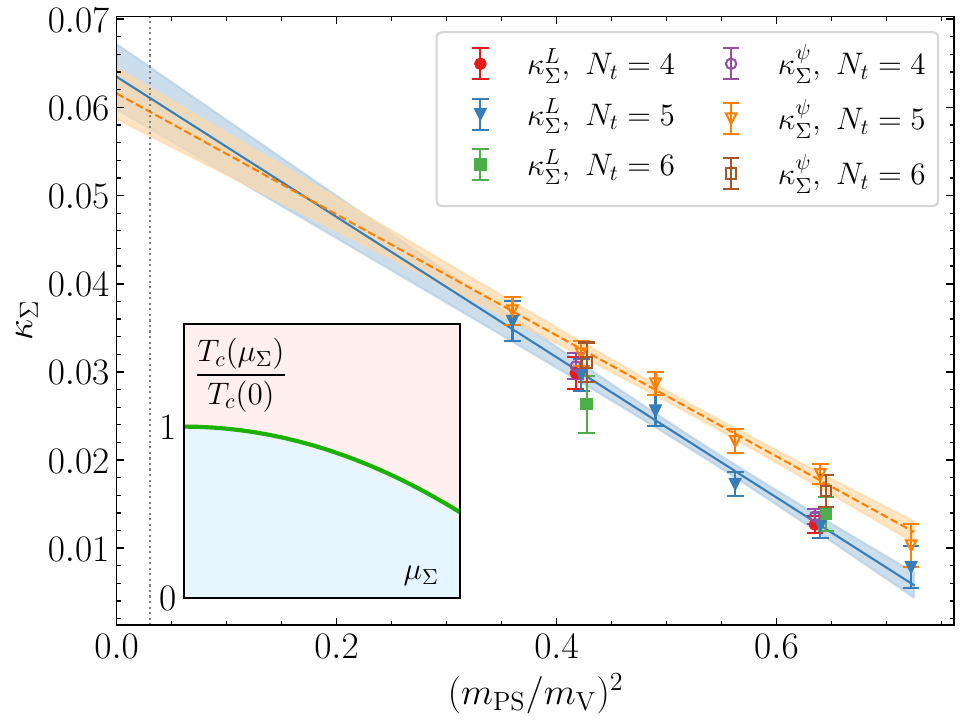} 
\caption{Curvatures of the deconfinement and chiral transitions, $\kappa_\Sigma^L$ and $\kappa_\Sigma^\psi$, respectively, which determine the parabolic dependence~\eqref{eq_Tc_mu_S} of the corresponding pseudocritical temperatures $T_c$ on the spin potential $\mu_\Sigma^{\rm I}$. The pseudocritical temperature can be well described by a quadratic function of function~\eqref{eq_Tc_ratio_fit} of the mesonic mass ratio $m_{\rm PS}/m_{\rm V}$. The solid/dotted lines represent the best fit of $N_t = 5$ data for deconfinement/chiral crossover by the parabolic function~\eqref{eq_Tc_ratio_fit} with the shadowed region denoting the error range of the fit. The vertical dashed line represents the physical value of the meson ratio, $m_{\rm PS}/m_{\rm V} \simeq 0.175$. The inset shows the qualitative behavior~\eqref{eq_Tc_mu_S} of the pseudocritical temperature $T_c$ on the spin potential $\mu_\Sigma$.}
\label{fig_kappa_S}
\end{figure}

The best fits of the data for the pseudocritical temperature by the quadratic polynomial~\eqref{eq_T_muI_S} of the imaginary spin potential are shown in Fig.~\ref{fig_Tc_muI} by solid lines. These fits allow us to obtain the dependence of the parabolic slope $\kappa_\Sigma$ as a function of the (squared) ratio of the meson masses, shown in Fig.~\ref{fig_kappa_S}. 

We immediately notice from Fig.~\ref{fig_kappa_S} that the curvature $\kappa_\Sigma$ of the spin-related shift in the pseudocritical temperature is a positive quantity. It implies, according to the analytical continuation~\eqref{eq_T_muI_S}, that the pseudocritical temperature decreases as the spin density becomes higher. 

Furthermore, we observe that as the pion mass increases (i.e., as the ratio $m_{\rm PS}/m_{\rm V}$ grows), the curvature $\kappa_\Sigma$ decrease. This behavior is anticipated, as the contributions of quark loop effects diminish with increasing quark mass. At sufficiently high masses, the impact of quarks on gluon dynamics, including the transfer of spin polarization from quarks to gluons, becomes negligible. This expectation is well corroborated by our data presented in Fig.~\ref{fig_Tc_muI}.

It appears that the dependence of the curvatures $\kappa^\ell_\Sigma$ ($\ell = L, \psi$) on the pion mass ratio can be well described by a simple parabolic dependence on the ratio of the meson masses:
\begin{align}
    \kappa^\ell_\Sigma(\xi) = \varkappa^\ell_\Sigma + \gamma^\ell_\Sigma \, \xi^2\,,
    \qquad
    \xi = \frac{m_{\rm PS}}{m_{\rm V}}\,.
	\label{eq_Tc_ratio_fit}
\end{align}
The best fits of the curvatures $\kappa^\ell_\Sigma$ with $\ell = L, \psi$ for $N_t = 5$ lattices by the empirical  function~\eqref{eq_Tc_ratio_fit} are shown in Fig.~\ref{fig_kappa_S}. The best-fit parameters are: 
\begin{align}
	\varkappa_\Sigma^L = 0.0635(37)\,,
    \qquad
    \gamma_\Sigma^L = -0.0795(67)\,, \\
	\varkappa_\Sigma^\psi = 0.0616(29)\,,
    \qquad
    \gamma_\Sigma^\psi = -0.0686(53)\,,
\end{align}
implying that at the physical ratio of the masses, the curvature of the spin-induced inhibition of the deconfinement temperature is:
\begin{subequations}
\begin{align}
	\kappa_\Sigma^{L\ {(\rm phys)}} & = 0.0610(35) \,, \\
	\kappa_\Sigma^{\psi\ {(\rm phys)}} & = 0.0595(27) \,,  \\
    & \qquad \left[ \ {\rm at} \ \biggl(\frac{m_{\rm PS}}{m_{\rm V}}\biggr)_{\rm phys} = 0.175\ \right] \,.
    \nonumber
\end{align}
    \label{eq_kappa_S_phys}
\end{subequations}

The influence of the finite spin density on the deconfining crossover is qualitatively similar to the effect produced by a non-vanishing baryon density, since a non-vanishing baryon chemical potential lowers the temperature of the deconfinement  crossover  as well. 

It is interesting to compare the curvatures in the crossover transition generated by finite spin and finite baryon densities. We can take, as a reasonable reference point, the curvature $\kappa_B$ of the pseudocritical crossover transition as a function $T_c(\mu_B)/T_c(0) = 1 - \kappa_B (\mu_B/T_c)^2$ of the baryon chemical potential $\mu_B$. Taking as an estimate the continuum-extrapolated result $\kappa_B = 0.0135(20)$ of Ref.~\cite{Bonati:2015bha} for QCD with $N_f = 2 + 1$ dynamical quarks and setting, for strange-neutral matter, the relation  $\mu_B = 3 \mu_q$ between the baryon, $\mu_B$, and flavor-averaged quark chemical potential $\mu_q$, we get $T_c(\mu_B)/T_c(0) = 1 - \kappa_q (\mu_q/T_c)^2$ with $\kappa_q = 9 \kappa_B$. Thus, the curvature of the pseudocritical temperature associated with the presence of a nonzero quark chemical potential $\mu_q$ is $\kappa_q = 0.12(2)$, which is approximately two times larger than the curvatures of the chiral and deconfinement  crossover ~\eqref{eq_kappa_S_phys}. Other numerical results for the curvature~\cite{Bonati:2018nut, Borsanyi:2020fev, Ding:2024sux} give similar estimations.
We conclude that the effect of a finite spin density on the crossover transition is quantitatively similar to the effect of the finite baryon density.

The effect of a finite isospin quark density on the transition temperature is also similar to the one of the finite quark density: the pseudocritical crossover temperature diminishes with the increase of the isospin chemical potential~\cite{Brandt:2017oyy}. The curvatures of both these crossovers are close to each other as well~\cite{Brandt:2016zdy}. Thus, we conclude that a finite spin polarization affects both thermal crossovers in a very similar way to the finite quark density and finite isospin density of quarks.

Despite the similarity mentioned above, there is an important difference between chiral and deconfinement thermal crossovers at finite baryon and spin densities. This difference can be seen from Fig.~\ref{fig_susceptibilities}: imaginary spin   potential noticeably increases the width of the chiral susceptibility, thus softening the chiral crossover transition, while there is no sign of similar softening  for the deconfinement crossover. Notice that this property is not observed at finite imaginary baryon chemical potential (see, for instance,  the recent paper \cite{Borsanyi:2025lim}). Analytically continuing the width of the chiral  crossover , one can draw a conclusion that the  crossover  becomes more abrupt for real values of spin   potential. It might be that at sufficiently large spin density the width of the  chiral  crossover  is zero and it might turn to the first-order phase transition. Notice, however, that a careful study of this point is beyond the scope of the present paper.

\section{Conclusions}

We performed first-principle numerical simulations in the lattice QCD with $N_f = 2$ dynamical quarks to determine the effect of a finite spin density of quarks on the deconfinement and chiral crossovers. The spin density has been introduced by employing a finite spin potential within the canonical definition of the Dirac fermion spin. 

Our simulations reveal that a finite spin density of quarks affects the dynamics of gluons and has a measurable impact on the confining and chiral properties of QCD. Thermodynamically, the quark spin density leads to a decrease in the temperatures of the chiral and deconfinement crossover~\eqref{eq_Tc_mu_S}, making the spin potential somewhat similar to the baryon chemical potential that has a qualitatively similar effect. This similarity can have a physical explanation since an increasing baryon chemical potential introduces quarks or anti-quarks into the system, favoring the deconfined phase by enhancing the screening of the long-range color potential, confining string breaking, and hadron percolation effects. Likewise, the spin potential introduces quarks and anti-quarks to produce a spin-polarized but globally baryon-neutral medium. The presence of the dynamical quark degrees of freedom in the latter case also leads to color screening and hadronic percolation, which inhibits the confining properties of gluons.

Quantitatively, the inhibition of the confining and chiral-symmetry-breaking properties of quark-gluon matter in the presence of a finite spin polarization of quarks can be described by the dimensionless curvatures, 
$\kappa_\Sigma^{L}$ and $\kappa_\Sigma^{\psi}$  associated with the deconfinement and chiral crossovers, respectively. We established the effects of the pion mass on the pseudocritical temperature and found that at the physical point, the  crossover  curvatures are approximately the same within a small statistical error~\eqref{eq_kappa_S_phys}, $\kappa_\Sigma^{L} \simeq \kappa_\Sigma^{\psi} \simeq 0.06$. The equivalence of the curvatures implies that the presence of the background spin potential does not lead to a splitting of the deconfinement and chiral crossovers.

The small magnitude of the curvatures implies that for the phenomenologically relevant values of the spin potential $\mu_\Sigma = 10\,{\rm MeV}$, the deconfinement and chiral transition temperatures drop only by about $0.03\%$. To make this estimation, we took into account that in rotating quark-gluon plasma, the magnitude of the spin polarization is expected to be of the order of vorticity, which reaches about $\Omega \simeq 10\,{\rm MeV}$ in experiments on heavy-ion collisions~\cite{STAR:2017ckg}. Our results imply that the spin polarization in the quark sector has a significantly smaller influence on the system properties than the rotation effects in the gluon sector~\cite{Braguta:2023iyx, Braguta:2024zpi, Braguta:2022str, Braguta:2023yjn, Braguta:2023kwl, Braguta:2023tqz}. Despite the tiny effect of a realistic spin density on the bulk thermodynamic  crossover , our result shows that the spin of quarks affects the deconfining properties of gluons, acting, presumably, via their spins. The existence of the transfer of spin polarization from quark to gluon degrees of freedom may be interesting on its own, as it may shed light on the well-known problem of the proton spin crisis~\cite{Ji:2020ena}.

\begin{acknowledgments}
The authors are grateful to Andrey Kotov for useful discussions. The work has been carried out using computing resources of the Federal collective usage center Complex for Simulation and Data Processing for Mega-science Facilities at NRC ``Kurchatov Institute'', \href{http://ckp.nrcki.ru/}{http://ckp.nrcki.ru/} and the Supercomputer ``Govorun'' of Joint Institute for Nuclear Research. The work of VVB and AAR, which consisted of the lattice calculation of the observables used in the paper and interpretation of the data,  was supported by the Russian Science Foundation (project no. 23-12-00072). The work of MNC was funded by the EU’s NextGenerationEU instrument through the National Recovery and Resilience Plan of Romania - Pillar III-C9-I8, managed by the Ministry of Research, Innovation and Digitization, within the project entitled ``Facets of Rotating Quark-Gluon Plasma'' (FORQ), contract no.~760079/23.05.2023 code CF 103/15.11.2022.  

\end{acknowledgments}

\bibliography{spin}

\end{document}